\documentclass{mn2e}
\usepackage{graphicx}
\usepackage{times}

\begin{document}

\title[GRB duration distribution]{On the observed duration
  distribution of gamma-ray bursts from collapsars}

\author[Lazzati et al.]{D. Lazzati$^1$, M.
  Villeneuve$^2$, D. L\'opez-C\'amara$^1$,
  B. J. Morsony$^3$, and R. Perna$^{2,4}$\\
$^1$Department of Physics, NC State University, 2401
  Stinson Drive, Raleigh, NC 27695-8202\\
$^2$Department of Astrophysical and Planetary Sciences,
  University of Colorado, Boulder, CO 80309-0391, USA\\
$^3$Department of Astronomy, University of
  Wisconsin-Madison, 3321 Sterling Hall, 475 N. Charter Street,
  Madison WI 53706-1582\\
$^4$JILA, University of Colorado and National Institute
  of Standards and Technology, Boulder, CO 80309-0440, USA}

\maketitle

\begin{abstract} 
  The duration of the prompt emission of long gamma-ray bursts is
  generally considered to be fairly similar to the duration of the
  activity of the engine in the center of the progenitor star. Here,
  we investigate the relation between the duration of the engine
  activity and that of the observed light curve, using inputs from
  both numerical simulations and observations. We find that the
  observed burst duration is a good proxy for the engine duration
  after the time necessary for the jet to break out the star's surface
  is subtracted. However, the observed duration is a function of the
  viewing angle and can be significantly shorter than the duration of
  the engine activity. We also show that the observed,
  redshift-corrected burst duration evolves only moderately with
  redshift for both observations and synthetic light curves. We
  conclude that the broad distribution of the observed duration of
  long BATSE gamma-ray bursts is mostly accounted for by an engine
  lasting $\sim20$~s, the dispersion being due to viewing and redshift
  effects. Our results do not rule out the existence of engines with
  very long duration. However, we find that they are constrained to be
  a small minority of the BATSE detected bursts.
\end{abstract}


\section{Introduction}

Gamma-ray bursts (GRBs), the most energetic explosive phenomena known
today, have historically been classified based on the time during
which 90\% of their flux is observed ($T_{90}$).  Since the early
epoch of the BATSE observations, it became apparent that the
distribution of durations is bimodal, with a clear minimum that,
depending on the sensitivity of the instrument, lies between 1 and 3
seconds (Kouveliotou et al. 1993; see also Nakar 2007).  Bursts with
durations between milliseconds to about 2 seconds are classified as
``short'' (SGRBs), while the class of ``long'' GRBs (LGRBs) is
characterized by a $T_{90}$ of more than 2 seconds, and the observed
distribution extends up to a few hundred seconds.

Based on the duration distribution and difference in the hardness
ratio, it was theorized that the progenitors of these bursts are of
different origin (Kouveliotou et al. 1993). SGRBs are now believed to
be associated with the merger of two compact objects (Nakar 2007),
while LGRBs are thought to be associated with the collapse of massive
stars (Woosley \& Bloom 2006). Even though alternative classification
schemes have been attempted (Gherels et al. 2006; Zhang et al. 2007;
Kann et al. 2011), the duration of a burst is critical to the
progenitor type.  Throughout the rest of this paper we will focus on
LGRBs.

Collapsars are Type Ibc core-collapse supernovae formed from massive
Wolf-Rayet stars that are rotating very fast and produce jets of
relativistic matter along their rotation axes (MacFadyen \& Woosley
1999, Nagataki 2011). The core of these progenitors evolves to become
a black hole, which accretes at super-Eddington rates from a disk
formed from the fallback material\footnote{An alternative model is the
  magnetar model in which the core evolves to a fastly spinning,
  highly magnetized neutron star (e.g. Usov 1992; Bucciantini et
  al. 2009). Since our jets are injected as boundary conditions, the
  true nature of the central engine is not important for the
  results.}. For a typical star, the jet takes about 10~s to cross the
star and break out on the surface (McFadyen \& Woosley 1999; Aloy et
al. 2000; Zhang et al. 2003; Morsony et al. 2007; Bromberg et
al. 2011, 2013; L\'opez-C\'amara et al. 2013).  This means that, for
GRBs with observed durations of $\sim 2$~s, the engine must be active
just a little longer than the breakout time.  On the other hand, at
the opposite end of the distribution of durations, engines active for
several hundreds of seconds are required.  Long engine durations are
invoked also to explain the observed X-ray flares, lasting up to
several hours (e.g. Falcone et al. 2007; Lazzati \& Perna 2007),
although in those cases the evolution of the accretion disk can
influence the observed duration (e.g. Perna et al. 2006).

The above considerations highlight the importance that the event
duration has for our understanding of the physics of the progenitor
star and the immediate aftermath of the explosion. The period of time
during which the engine that powers the GRB jet is active is known as
$T_{\rm{eng}}$. The burst's $T_{90}$ has been generally considered to
be a close proxy for its $T_{\rm{eng}}$, at least after the jet has
broken out of the stellar surface (Kobayashi et al. 1997; MacFadyen \&
Woosley 1999; Bromberg et al. 2011, but see Lazzati et al. 2010). This
crucial assumption hence directly relates the observed $T_{90}$ to the
processes in the core of the progenitor star. The aim of this paper is
to test the validity of this assumption, and explore various factors
that could influence the observed duration of the burst. We achieve
this goal by means of two different and complementary
analyses. Firstly, we consider a sample of GRBs with well measured
redshifts and light curves. We derive the intrinsic light curves and
then place them within a redshift range, to examine the $z$-dependent
variation of the inferred observed duration (instrumental effects and
the gamma-ray background playing an important role). Second, and more
generally, we generate a large set of synthetic GRB light-power curves
from 2D relativistic hydrodynamic simulations of collapsars. We
simulate engines of different durations, and explore how the observed
GRB $T_{90}$ varies with viewing angle and redshift of the burst. We
are hence able to directly relate the intrinsic length of the GRB
engine to what is observed. Last, we create a mock catalogue of GRBs,
randomly distributed on the sky with viewing angles, and with a
redshift distribution that follows the star formation rate. We explore
the effect of different engine durations on the observed GRB
distribution, in an attempt to constrain the physical properties of
the GRB progenitors.

Our paper is organized as follows. Sect.~2 describes the simulations
that we use. The computation of the synthetic lights curves, as well
as the properties of the observed and simulated bursts, are described
in Sect.~3.  We then discuss the results and summarize in
Sect.~4. Throughout this paper, we adopt a standard $\Lambda$CBM
cosmology with $h_0=0.71$, $\Omega_M=0.27$, and $\Omega_\Lambda=0.73$.

\begin{figure}
\includegraphics[width=\columnwidth]{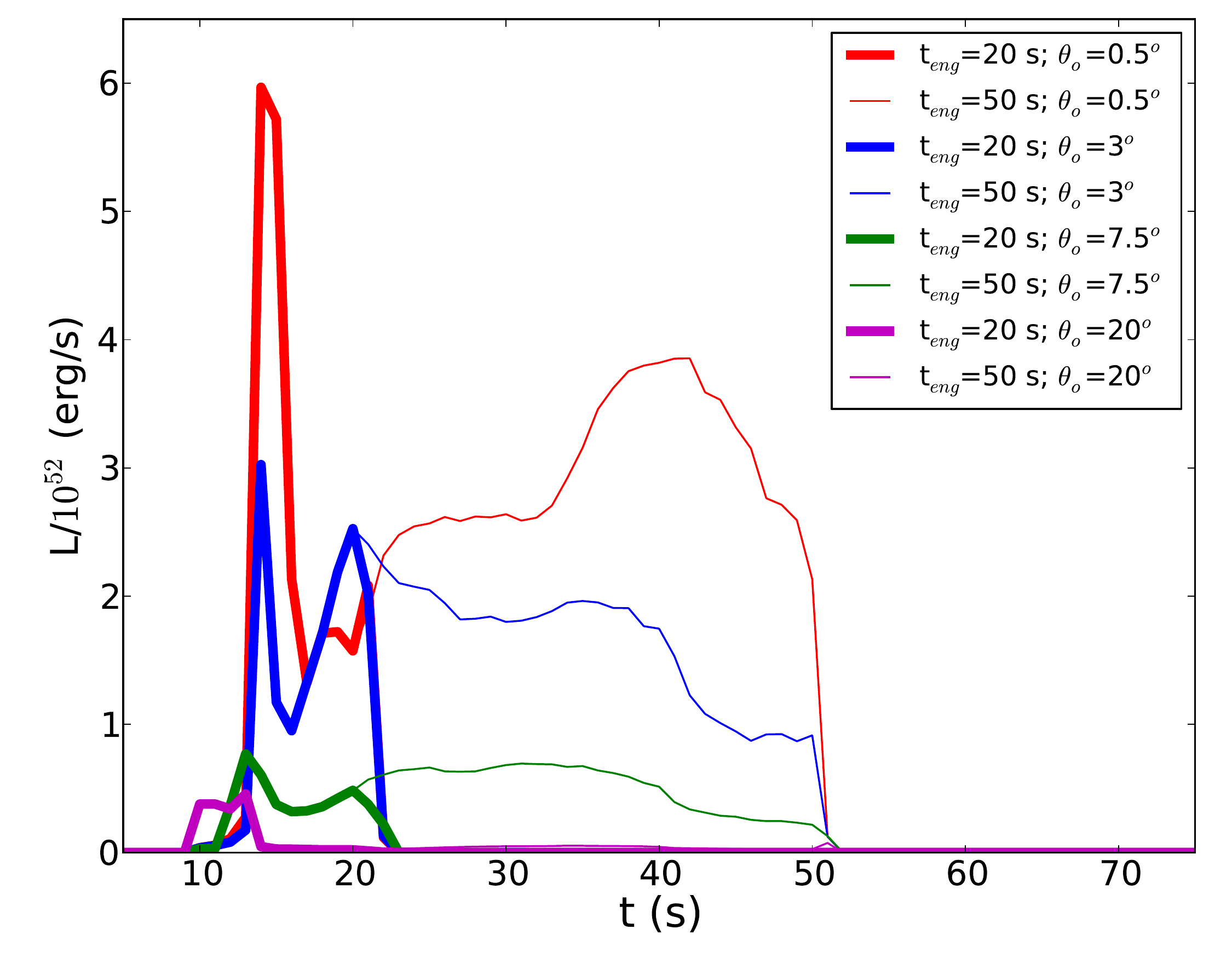}
\caption{{Bolometric light-power curves extracted from the simulations
    with $T_{\rm{eng}}=20$ and 50 s at various off-axis angles. Note
    how the light-power curves when the engine is active are not
    affected by the engine duration: the 20~s light-power curve
    coincides with the 50~s light-power curve for $t<20$~s.}
\label{fig:lc}}
\end{figure}

\section{Numerical simulations}

All the simulations presented in this paper were performed with the
special-relativistic, adaptive-mesh-refinement hydrodynamic code
FLASH, version 2.5 (Fryxell et al. 2000). We adopted a 9 level AMR
mesh with a maximum resolution of $3.9\times10^6$~cm at the highest
level of refinement. At this resolution the transverse dimension of
the injected jet was resolved into at least 46 elements. Our
simulations did not include magnetic fields, due to the technical
challenge of performing MHD calculations with relativistic motions on
an adaptive mesh. In addition, gravity from a central mass and
self-gravity were neglected since the characteristic times of the
jet-star interaction are much shorter than the dynamical time of the
progenitor star's collapse (see, e.g., Lazzati et al. 2012).

\begin{figure}
\includegraphics[width=\columnwidth]{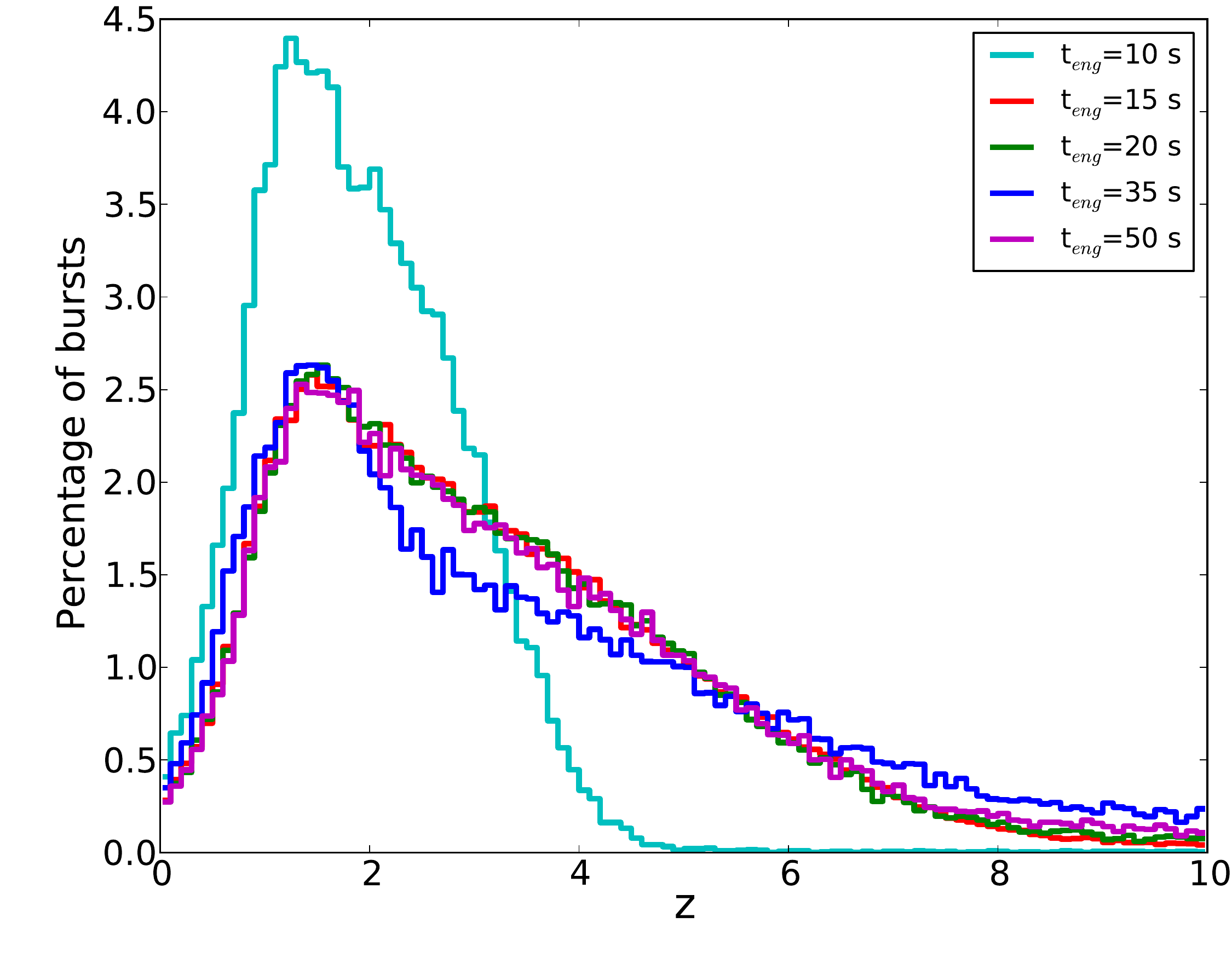}
\caption{{Redshift distribution of the detected bursts from the mock
    catalogs for each of the simulated engines. The vertical axis
    shows the percentage of bursts falling in each redshift bin.}
\label{fig:z}}
\end{figure}
 
All our simulations adopted a realistic GRB stellar
progenitor. Specifically, we used model 16TI from Woosley \& Heger's
(2006) one dimensional pre-SN models. 16TI is a 16 solar-mass
Wolf-Rayet star which is rapidly rotating
($J_{0}=3.3\times10^{52}$~erg~s), and has low metallicity (1\% solar)
at the ZAMS. The mass of the star at pre-explosion is 13.95 solar
masses and its radius is $4.07\times10^{10}$~cm. The progenitor, which
was mapped into a two dimensional domain (assuming cylindrical
symmetry), was placed in a constant density ISM ($\rho_{ISM} =
10^{-10}$~g~cm$^{-3}$), and had a relativistic jet injected at the
inner boundary ($R_0=10^9$~cm).  The jet, with an initial Lorentz
factor of $\Gamma_0=5$, a half-opening angle $\theta_j=10^\circ$, and
with enough internal energy to reach an asymptotic Lorentz factor
$\Gamma_\infty=400$ upon complete non-dissipative acceleration, was
injected with a luminosity equal to 5.33$\times$10$^{50}$~erg~s$^{-1}$
for a certain amount of time (depending on the model). Five different
models were used, each with different engine duration
($T_{\rm{eng}}$). The durations were 10, 15, 20, 35, and 50~s, after
which the jet was turned off for the rest of the 100~s total
computation time. The equatorial boundary was set to reflective for
the whole duration of the simulation, while the two outer boundaries
were set at all times with free outflow conditions. The simulation box
was $2.56\times10^{11}$~cm in length (along the jet direction) and
$1.28\times10^{11}$~cm across.


\section{Synthetic Light Curves}

\begin{figure}
\includegraphics[width=\columnwidth]{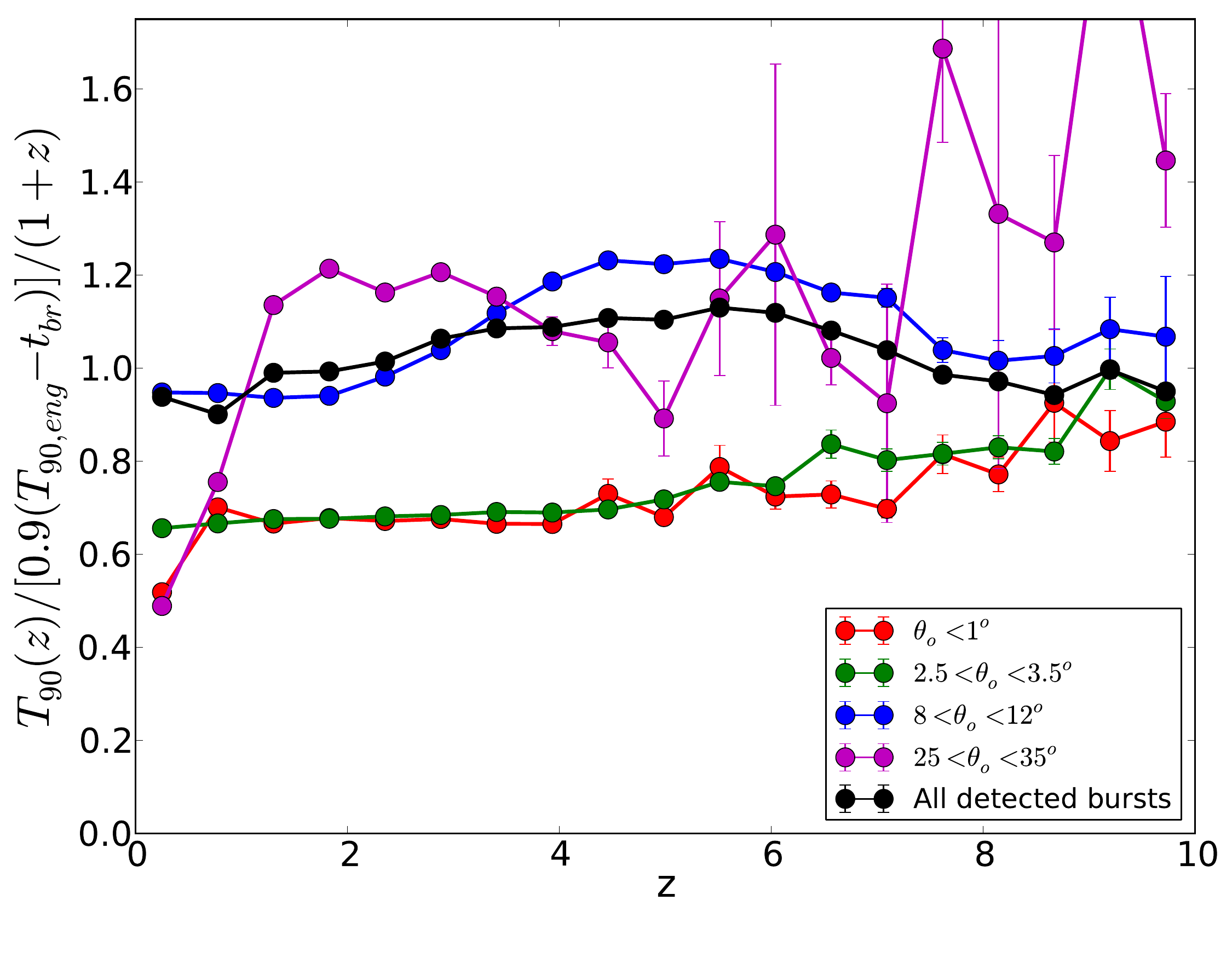}
\caption{{Ratio of the duration of the simulated detector light-power
    curve over the engine 90 percentile duration (after subtracting
    the breakout time) normalized for cosmological time dilation for
    the 20~s engine simulation. Different curves show different lines
    of sight to the observer, as detailed in the legend. Solid lines
    show the average over many simulations, while error bars, where
    visible, show the uncertainty associated to the limited size of
    the simulated sample. All points have an error bar associated with
    them, but in most cases it is smaller than the marker.}
\label{fig:t90_sim}}
\end{figure}
 
Light-power curves for the four engine durations were extracted from
the 2D simulations following Morsony et al. (2010) for viewing angles
between 0.5 and 90 degrees off-axis, assuming 50\% radiative
efficiency (Figure~\ref{fig:lc}). The figure shows that the
light-power curves do not depend on the engine duration while the
engine is active. This is not surprising since the engine luminosity
is the same for every simulation, and the jet head is not in causal
connection with its base. If we had chosen to keep the total ejected
energy constant and made the shorter engines more luminous, the light
curves from short and long engines would not have overlapped, even at
the early times when both engines are active.  Bolometric light-power
curves were converted into observed countrates as:
\begin{equation}
\Phi=B+\frac{AL_{\rm{bol}}}{4\pi d_L^2}
\int_{\epsilon_m}^{\epsilon_M}
\frac{B_{\alpha,\beta,E_0/(1+z)}(\epsilon)}{1.6\times10^{-9}\epsilon}d\epsilon\;,
\label{eq}
\end{equation}
where $\Phi$ is in counts per second, $B$ is the background in counts
per second, $A$ is the effective area of the adopted instrument in
cm$^2$, $\epsilon_m$ and $\epsilon_M$ are the lower and upper
sensitivity bounds of the adopted instrument in keV, respectively,
$L_{\rm{bol}}$ is the bolometric luminosity in erg/s, $d_L$ is the
luminosity distance in cm, $\epsilon$ is the photon energy in keV, and
$B_{\alpha,\beta,E_0}(\epsilon)$ is a normalized Band function (Band
1983) with low-frequency slope $\alpha$, high-frequency slope $\beta$,
and peak frequency $E_0$. The Band function is normalized so that:
\begin{equation}
\int_0^\infty B_{\alpha,\beta,E_0}(\epsilon)\,d\epsilon=1
\end{equation}
In order to mimic GRBs detected by BATSE, we assumed $B=10000$,
$[\epsilon_m,\epsilon_M]=[25,2000]$~keV, and $A=2000$~cm$^2$. Average
values for the spectral slopes $\alpha=0$ and $\beta=-1.5$ were also
assumed\footnote{Note that we here define $\alpha$ and $\beta$ as the
  spectral indices, not the photon indices.}. Finally, the peak
frequency $E_0$ was set to follow the intra-burst
luminosity-peak-frequency correlation and obey (Ghirlanda et al 2010;
Lu et al 2012):
\begin{equation}
E_0=10^{0.53\log_{10}(L_{\rm{bol}})-25.3}\;.
\end{equation}

Mock GRB catalogs were constructed for each engine duration by
exploding massive stars at a rate proportional to star
formation\footnote{We note that the assumption that the GRB rate
  strictly follows the star formation rate is highly debated in the
  literature. Several studies (e.g. Wanderman \& Piran 2010) suggest
  that the relative GRB-to-star formation rate is higher at higher
  redshift, although observational biases might contribute to this
  effect (e.g. Kistler et al. 2009; Robertson \& Ellis 2012; Trenti et
  al. 2012).} SFR2 in Porciani \& Madau (2001). SFR2 assumes that the
comoving star formation rate is flat at high redshift. The exploding
stars were randomly oriented with respect to the line of sight and a
detection algorithm was applied to the resulting synthetic countrate
curves. Countrate curves that produced a 5-$\sigma$ increase in the
countrate on at least one timescale among 1, 5, and 25 seconds were
labeled as ``detections''. Figure~\ref{fig:z} shows the redshift
distribution of the detected busts for the five different engine
durations.

\begin{figure}
\includegraphics[width=\columnwidth]{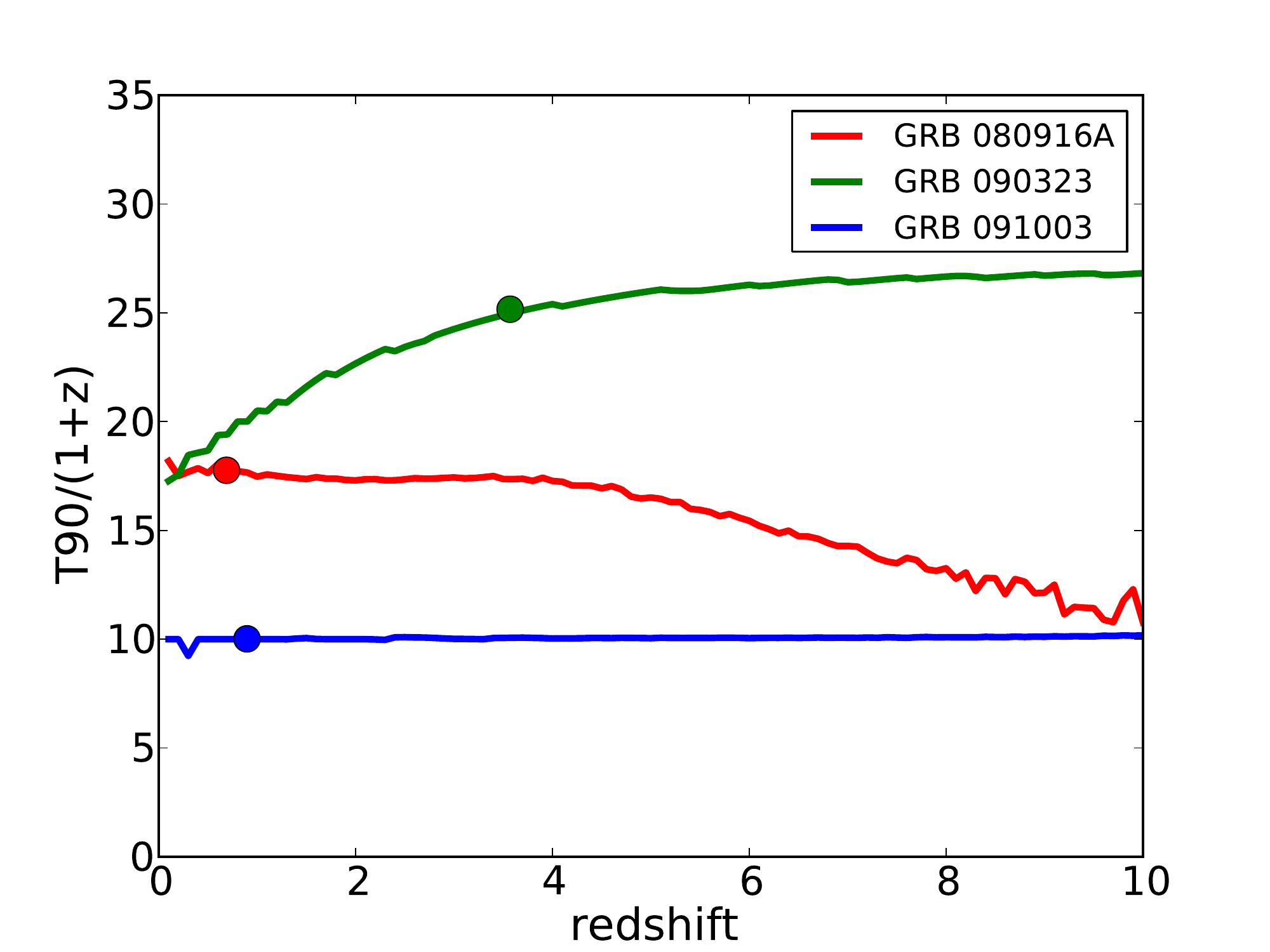}
\caption{{Detected duration of selected Fermi GRBs as a function of
    hypothetic redshift. A solid symbol shows, for each burst, the
    redshift and turation with which it was actually detected.}
\label{fig:t90_obs}}
\end{figure}
 
Figure~\ref{fig:t90_sim} shows the redshift dependence of the ratio of
the $T_{90}(z)$ measured from the synthetic light-power curves over
the expected duration expressed as $0.9(T_{\rm{eng}}-T_{\rm{br}})$,
i.e., 90\% of the time interval during which the engine energy is
released minus the breakout time. In our simulations the engine had
constant luminosity and $T_{\rm{bo}}$ was measured to be
$\sim8$~s. The curves were also divided by $(1+z)$ to remove the
cosmological dilation and enhance the role of spectral evolution on
the measurement of the duration of the light curve. The figure shows
results for the $T_{\rm{eng}}=20$~s simulation, but results from the
other simulations are similar, except for the $T_{\rm{eng}}=10$~s
simulation, where very few bursts are detected at large redshifts,
even for small viewing angle (see Figure~\ref{fig:z}). This seems to
be due to the fact that these shorter bursts have lower fluence and
are not detectable in the longer time-scale of 25 seconds.

\begin{figure*}
\parbox{\columnwidth}{\includegraphics[width=\columnwidth]{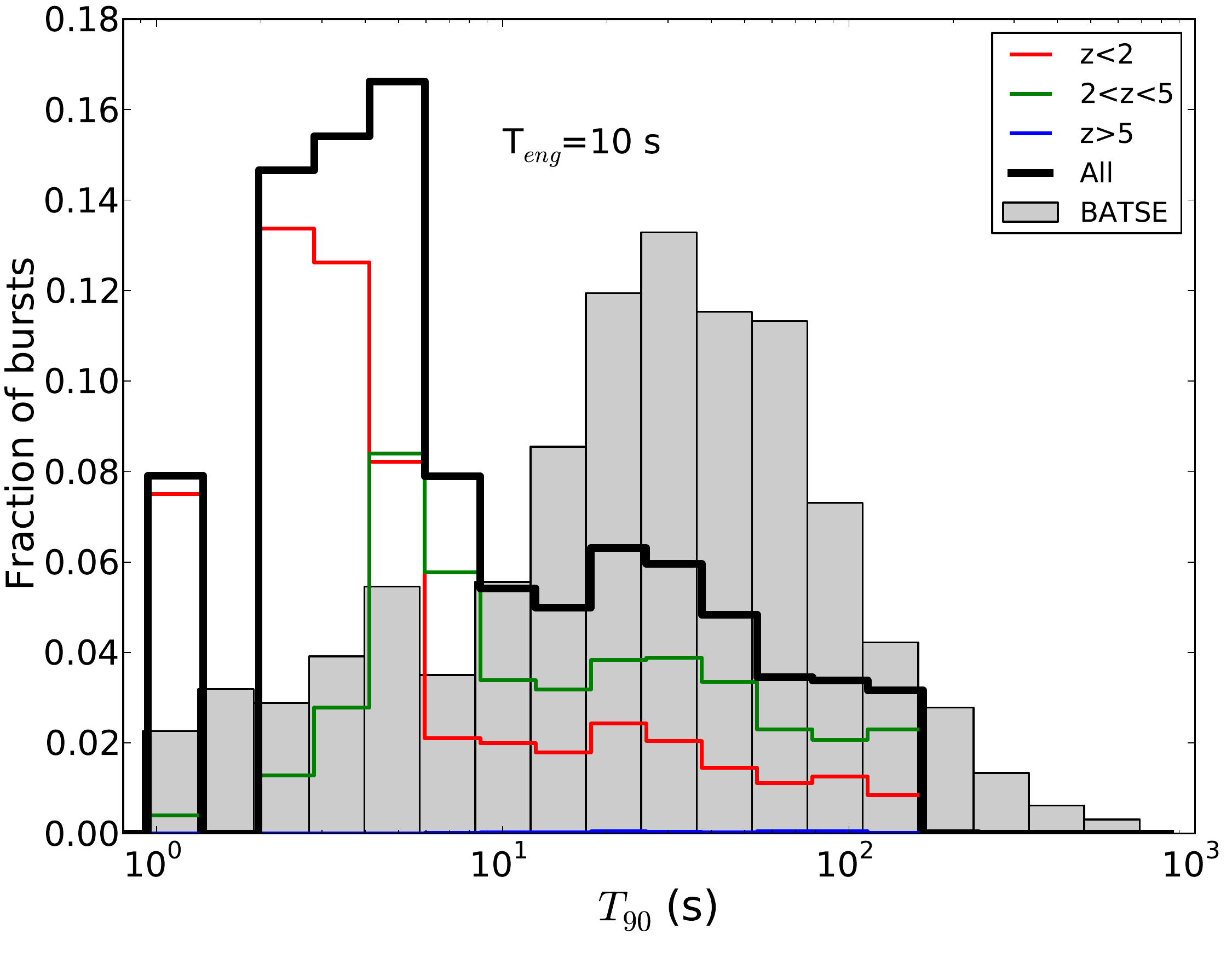}}
\hspace{2mm}
\parbox{\columnwidth}{\includegraphics[width=\columnwidth]{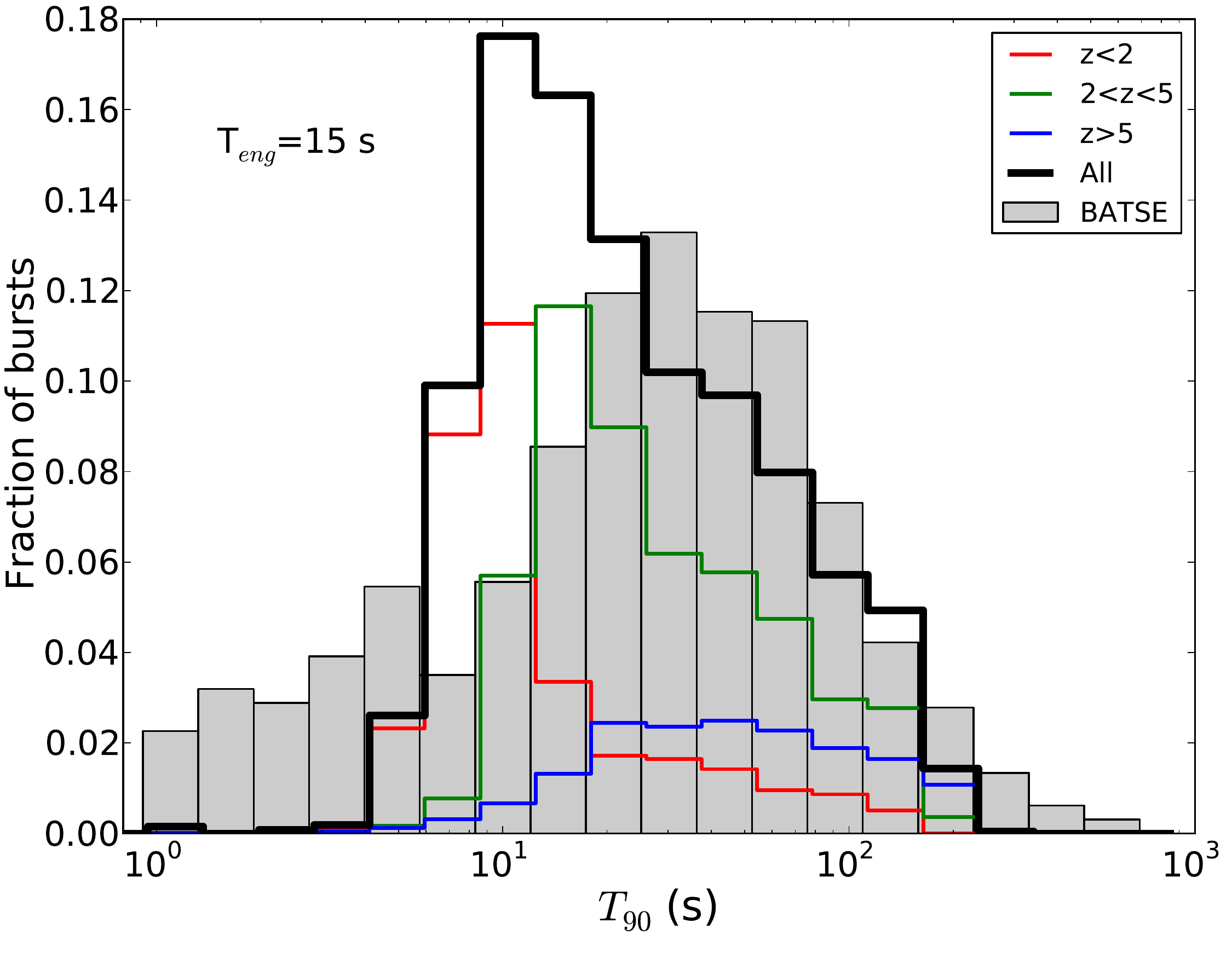}}
\parbox{\columnwidth}{\includegraphics[width=\columnwidth]{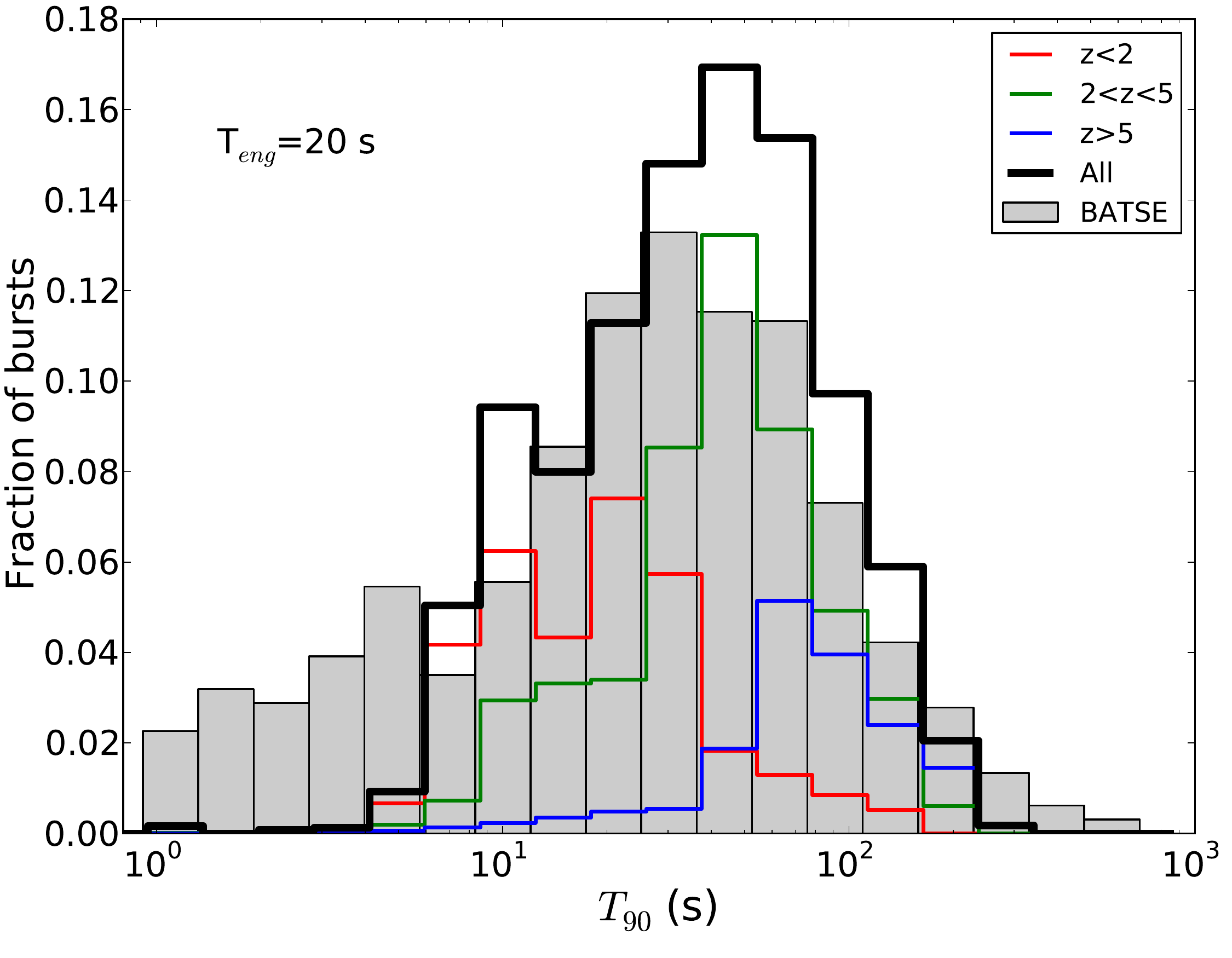}}
\hspace{2mm}
\parbox{\columnwidth}{\includegraphics[width=\columnwidth]{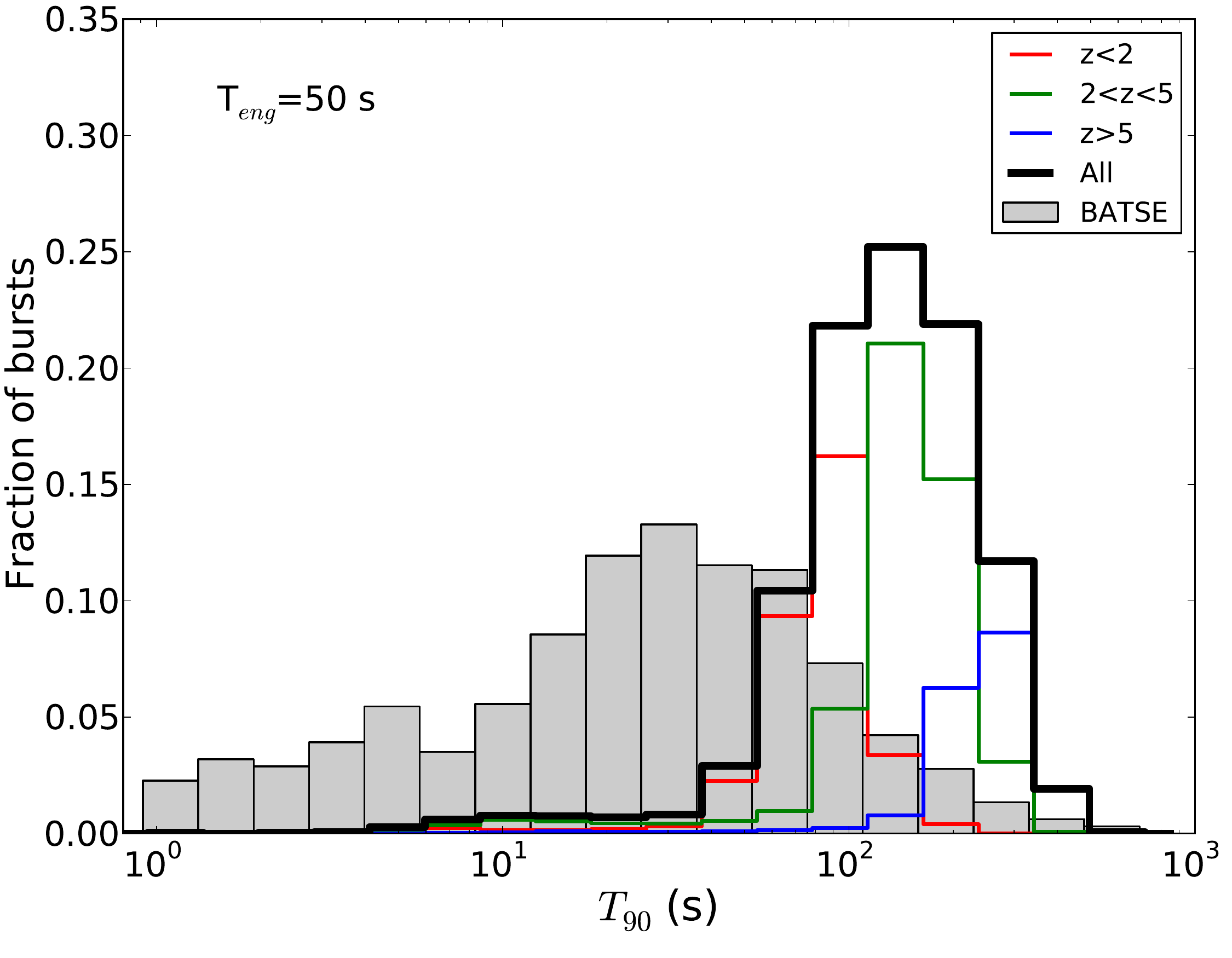}}
\caption{{Distribution of the BATSE duration catalog (shaded
    histogram) and of the mock catalogs generated from simulations
    with $T_{\rm{eng}}=15$~s (upper left panel), 20~s (upper right
    panel), 35~s (lower left panel), and 50~s (lower right panel). In
    all panels, the thick solid line is the duration distribution of
    the whole mock catalog, while the colored lines show the partial
    distribution in selected redshift bins.}
\label{fig:t90}}
\end{figure*}
  
Figure~\ref{fig:t90_sim} illustrates two important findings: first,
the observed light-power curve duration is on average a good proxy for
the engine duration, if the breakout time is subtracted. However, the
$T_{90}$ can be much shorter than the engine duration, especially for
bright, on-axis, low-redshift events. This is partly due to the fact
that hydrodynamic collimation makes the jet opening angle a
time-dependent quantity, causing some observers to see only very dim
emission when the jet angle is smaller than their viewing angle.
Indeed, the $T_{90}$ duration measured from the light curve has a
marked dependence on the viewing angle, the on-axis bursts lasting
less the off-axis ones for the same jet-progenitor pair. Second, there
is only modest evolution of the observed $T_{90}$ with redshift, once
the dominant $(1+z)$ component is removed. The figure suggests a
moderate increase of the duration with redshift for on-axis observers,
a decrease for wide-angle observers, and a mixed behavior for
observers at intermediate angles. At large redshifts, only a few of
the simulated curves are detected, and the measured duration is
dominated by the statistical error, creating the impression of a
growth of the duration with redshift which is instead due to the fact
that the burst is so faint that large chunks of background are
mistakenly included in the $T_{90}$ measurement. This inaccuracy is
intrinsic to non-imaging counting instruments like BATSE, and we
haven't therefore applied any correction since we are comparing to
BATSE results. In the same figure, a black line shows the average
$T_{90}$ observed as a function of redshift for the detected
bursts. The black line takes into account the fact that there are less
small viewing angle bursts and that large viewing angle bursts are
intrinsically faint and hard to detect. For these reasons, it mimics
fairly closely the intermediate viewing angle line.

In order to compare the results of Figure~\ref{fig:t90_sim} with
observations, we selected three GRBs observed by Fermi in a wide
spectral range for which extensive time-dependent spectral analysis
has been performed (Lu et al. 2012). These are GRB~080916A,
GRB~090323, and GRB~091003, whose general properties ($\alpha$,
$\beta$, $E_{\rm{pk}}$ and flux, from Lu et al. 2012) are reported in
Table~\ref{tab:grbs}.  Then, Eq.~\ref{eq} was used to convert such
properties into simulated light-power curves for the same burst as
they would have been observed from different redshifts. The results of
these calculations are seen in Figure~\ref{fig:t90_obs}. As for the
simulated light-power curves, we see that there is only a moderate
evolution with redshift, once the effect of cosmological time dilation
has been removed. One burst has a constant observed $T_{90}$, one has
a growing one, and one has a decreasing one. The different behavior is
connected to the peak frequency of the emission. Softer bursts are
consistently redshifted out of the instrument band and become
therefore shorter with redshift. Harder bursts are instead redshifted
into the instrument band and become longer with redshift.

\begin{table}
\begin{center}
	\begin{tabular}{|c|c|c|c|c|}
	\hline
	GRB & \emph{z}  & $E_{pk} [keV]$ & Fluence[erg/cm$^2$]\\
	\hline \hline
	080916A & 0.689 & 107.41 & 7.81e-6\\
	\hline
	090323 & 3.5774 & 639.31 & 1.18e-4\\
	\hline
	091003 & 0.8969 & 366.44 & 2.33e-5\\
	\hline 
\end{tabular}
\end{center}
\caption{{Properties of the observed GRBs used for computing the
    redshift evolution of their observed $T_{90}$ duration.}
\label{tab:grbs}}
\end{table}

\begin{figure}
\includegraphics[width=\columnwidth]{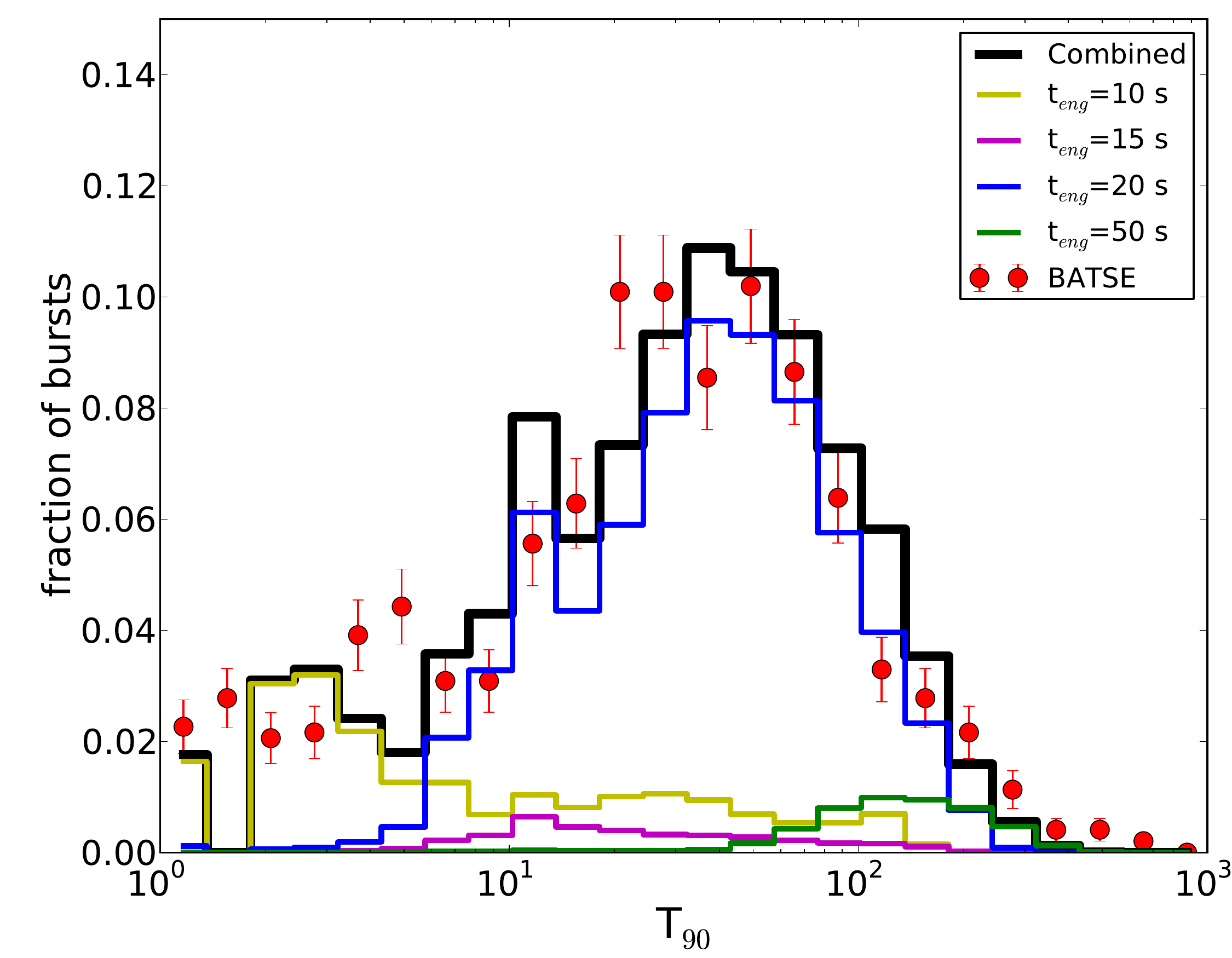}
\caption{{Fit of the BATSE duration distribution of long GRBs with a
    linear combination of the mock catalogs for different engine durations.}
\label{fig:fit}}
\end{figure}

A final comparison can be performed between the BATSE $T_{90}$
distribution and the distribution of the $T_{90}$ of the simulated
bursts from the mock catalogs. Each panel in Figure~\ref{fig:t90}
shows the comparison for a different engine duration. The BATSE
distribution is always shown with a shaded histogram, while the mock
distribution is shown with a thick solid line. For the mock catalogs
only, the distribution of durations from bursts in selected redshift
bins is shown with colored lines. The comparison shows that most of
the width of the duration distribution can be accounted for by a
single progenitor star and engine duration combination. Several
effects contribute in broadening the duration distribution from a
single progenitor. First, the redshift time-dilation systematically
increases the observed duration of distant bursts. Second, the
uncertainty associated with the difficulty of selecting the starting
and ending point of a weak GRB overlaid on a noisy background can
affect both positively and negatively the duration of a burst,
especially if observed at wide angle and/or at high redshift. This
effect is the main contributor to the population of bursts with
detected durations much shorter than the engine duration. Finally, the
dependence of the observed duration on the viewing angle can
moderately decrease the observed duration (up to a factor $\sim2$ for
the brightest bursts).  To quantify this statement, we show in
Figure~\ref{fig:fit} the result of fitting the BATSE distribution with
a linear combination of the distributions from the five mock
catalogs. The BATSE distribution was purged of any burts with
$T_{90}<1$~s in order to eliminate short-duration GRBs that cannot be
accounted for in this model. The fit was performed by minimizing the
$\chi^2$ allowing for the BATSE detected burst to have free
contributions from the various engine durations (each population
percentage being allowed to vary independently between 0 and 100 and
constraining the sum of all percentages to be equal to 100). We find
that the overall shape of the BATSE duration distribution can be
accounted for by an engine duration distribution strongly peaked at
$T_{\rm{eng}}=20$~s (70\% of the bursts) with a small component of
shorter engines (21\% with $T_{\rm{eng}}=10$~s and 4\% with
$T_{\rm{eng}}=15$~s) and a minor component of longer engines (5\% with
$T_{\rm{eng}}=50$~s). No engine with $T_{\rm{eng}}=35$~s is required
to fit the BATSE distribution.

\section{Discussion and Conclusions}

We have used numerical simulations to investigate the connection
between the intrinsic engine duration and the light curve duration in long GRBs
and the origin of the observed duration distribution. Our main
conclusions and a discussion of their caveats, limitations, and
implications are:

\begin{enumerate}
\item The observed light curve duration of the prompt emission of GRBs
  is, on average, a good proxy for the engine duration, after
  subtracting the jet breakout time. However, our study reveals a
  dependence of the burst duration on the viewing angle especially for
  burst at low redshift observed very close to the jet axis. The same
  progenitor and engine pair can therefore produce somewhat different
  bursts, those observed on-axis being brighter and shorter and those
  observed at an off-axis angle being longer and fainter.

\item The duration of the same event hypothetically seen at different
  redshifts is fairly constant, once the effect of cosmological time
  dilation is accounted for, in both simulated and observed bursts. In
  this regard it is important to point out that our simulated engines
  had constant luminosity until they were suddenly shut-off. Engines
  with smoothly decaying tails could produce bursts with long-lasting
  X-ray tails that may produce a stronger effect when viewed at
  different redshifts. In the case of observed light curves, our
  method could have missed X-ray tails as well, since we relied on the
  possibility of performing spectral analysis of the observed light
  curves and therefore of a minimum number of photon counts. Soft,
  weak, long-lasting X-ray tails would have been missed in the
  observations as well as in the simulations. The consistent
  qualitative behavior between simulated and observed light curves
  should therefore be seen more as an indication that they were
  analyzed consistently rather than as a strong evidence of the lack
  of a redshift-dependent duration evolution.

\item Assuming a flat star formation rate at high-redshift, most of
  the BATSE duration distributions of long GRBs can be accounted for
  with only a very small range of engine durations (between 10 and
  20~s). Only a very small fraction of very long-lasting engines are
  required to explain the long duration tail and, possibly, the
  recently-detected population of very long bursts (Gendre et
  al. 2013; Levan et al. 2013). The width of the BATSE duration
  distribution is almost entirely accounted for by the effect of the
  viewing angle and of the broad range of redshift at which bursts can
  be detected. The observed burst population is therefore dominated by
  engines with a duration significantly smaller than what has been
  explored numerically (e.g. Lazzati et al. 2010, 2013). We show
  instead with this study that the typical engine duration of a burst
  from a compact Wolf-Rayet star is $\sim20$~s.
\end{enumerate}

\section*{Acknowledgements}
We thank S.E. Woosley and A. Heger for making their pre-SN models
available. The software used in this work was in part developed by the
DOE-supported ASC/Alliance Center for Astrophysical Thermonuclear
Flashes at the University of Chicago. MV acknowledges support from NSF
grant AST-1062736 for her participation in the URCA summer program at
NCSU, where this work was initially developed, and the Undergraduate
Research Opportunities Program at the University of Colorado
Boulder. This work was supported in part by the Fermi GI program
grants NNX10AP55G and NNX12AO74G (DL \& DLC), and by NSF Grant No. AST
1009396 (RP).  BJM is supported by an NSF Astronomy and Astrophysics
Postdoctoral Fellowship under award AST-1102796.

\end{document}